\documentclass{pasj00}

\begin{document}
\SetRunningHead{M. Uemura et al.}{}
\Received{2003/01/29}
\Accepted{2003/03/05}

\title{Discovery of a short plateau phase in the early evolution of a
gamma-ray burst afterglow}

\author{Makoto \textsc{Uemura}, Taichi \textsc{Kato}, Ryoko
\textsc{Ishioka}}
\affil{Department of Astronomy, Faculty of Science, Kyoto University,
Sakyou-ku, Kyoto 606-8502}
\email{uemura@kusastro.kyoto-u.ac.jp}
\and
\author{Hitoshi \textsc{Yamaoka}}
\affil{Faculty of Science, Kyushu University, Fukuoka 810-8560}

%

\KeyWords{gamma rays: bursts} 

\maketitle

\begin{abstract}
 We report optical observations during the first hour of the gamma-ray
 burst (GRB) afterglow of GRB021004.  Our observation revealed the
 existence of a short plateau phase, in which the afterglow remained at
 almost constant brightness, before an ordinary rapid fading phase.
 This plateau phase lasted for about 2 hours from 0.024 to 0.10 d after
 the burst, which corresponds to a missing blank of the early afterglow
 light curve of GRB990123.  We propose that the plateau phase can be
 interpreted as the natural evolution of synchrotron emission from the
 forward shock region of a blast wave.  The time when the typical
 frequency of the synchrotron emission passes through the optical range
 has been predicted to be about 0.1 d after the burst, which is
 consistent with the observed light curve.  Our scenario hence implies
 that the observed feature in GRB021004 is a common nature of GRB
 afterglows. 
\end{abstract}

\section{Introduction}
Gamma-ray bursts (GRBs) are the most energetic phenomena in the universe, 
which appear only for 0.01--100 s as brightest sources at the gamma-ray 
range (\cite{kle73grboriginal}; \cite{wij97GRB970228}).  In a
number of GRBs, afterglows have been detected in all wavelengths
(\cite{cos97xrayafterglow}; \cite{fra97radioafterglow}).  While their
central engine is still poorly understood, it has been proposed that
their behaviour can be well explained with a picture called the fireball
model.  In this model, a blast wave produced from a fireball propagates
outward initially at highly relativistic velocity, and then, gradually
decelerated by the collision with interstellar medium
(\cite{wax97fireball}).  The GRB afterglow is proposed to be synchrotron
emission from a forward shock region in an expanding shell colliding
with external medium, whereas the GRB itself is from an internal shock
region between shells (\cite{wij97GRB970228}; \cite{wax97afterglow};
\cite{sar99prediction}).  Early phase observations of afterglows would
provide crucial clues for the initial extremely relativistic state of
GRBs, however their rapid fading has prevented us from successful
observations. 

GRB021004 was detected by the HETE-2 satellite on 4 October 2002 at
12:06:13.57 UT (\cite{shi02gcn1565}).  A prompt identification of its
burst position, which was notified 48 s after the burst, enabled
observers to make observations of a very early phase of the afterglow of
the GRB. A new bright optical source of $R_{\rm c}=15.34$ was discovered
$6.56\times 10^{-3}$ day after the burst within the errorbox of the
burst position, and then continuously monitored by a number of observers 
(\cite{fox02gcn1564}).  Owing to the quick notification and the
apparently bright afterglow, GRB021004 provided a dense optical sampling
of the time evolution of the GRB afterglow.  Optical spectroscopic
observations showed its redshift of $z=2.32$ and its isotropic energy of 
$E_{\rm iso}=5.6\times 10^{52}\;{\rm erg\, s^{-1}}$
(\cite{mal02gcn1607}).  

\section{Observation and Result}
\begin{figure}
  \begin{center}
    \FigureFile(85mm,85mm){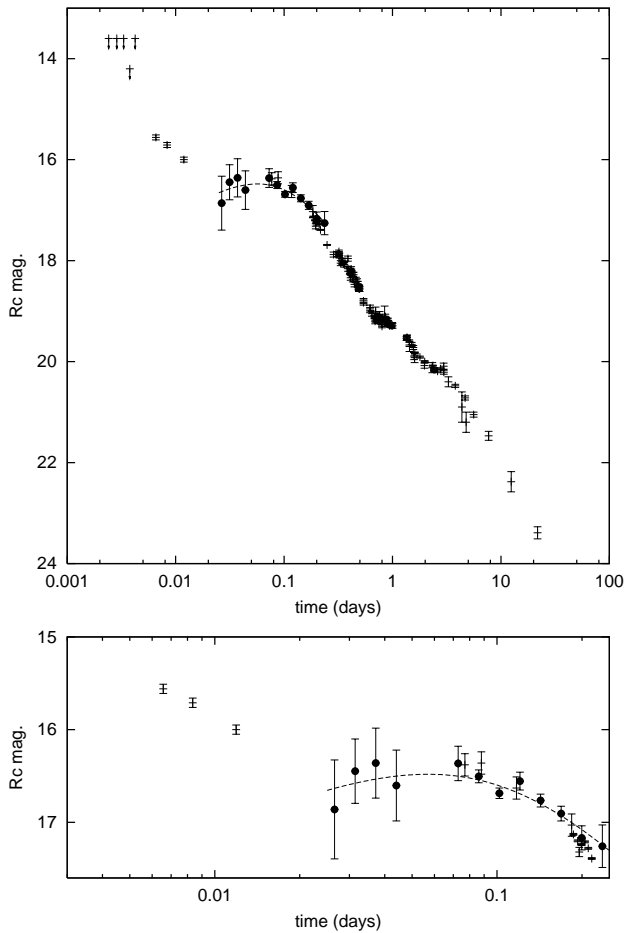}
  \end{center}
  \caption{Light curve of the optical afterglow of GRB021004.  The
abscissa and ordinate denote the time after the burst in day and the
 $R_{\rm c}$-magnitude, respectively.  Our observations are shown in the
 filled  circle.  The other symbols are observations reported in  GCN
 1564, 1573, 1576, 1577, 1578, 1580, 1582, 1584, 1585, 1586, 1587, 1591,
 1593, 1594, 1597, 1598, 1602, 1603, 1606, 1607, 1614, 1615, 1616, 1618,
 1623, 1628, 1638, 1645, 1652, 1654, and 1661.  The reported magnitudes
 in GCN were translated using the standard sequence presented in GCN
 1630.  The dashed line is the best fitted model light curve (see the
 text). Upper panel: The whole light curve of the afterglow.  Lower
 panel: Enlarged light curve around our observations.}
\label{fig:lc}
\end{figure}

Our unfiltered CCD photometric observations were performed with 25-cm
and 30-cm telescopes at Kyoto University.  The dark current image was 
subtracted from obtained CCD images, and then flat fielding was
performed.  We calculated differential magnitudes of GRB021004 with
neighbour comparison stars (GSC1183.403, GSC1183.1526, and
GSC1183.1261), whose constancy is better than 0.08 mag during our
observation.  We discarded images taken under heavily low transparency
conditions to avoid systematic errors.  Our equipment yields a magnitude
system near the $R_{\rm c}$-system since the sensitivity peak of the
camera is near the peak of the $R_{\rm c}$-system and spectra of GRB
afterglows have smooth continuum without strong emission lines or
absorption edge in the optical range.  The difference between our
unfiltered CCD and the $R_{\rm c}$-system is less than 0.02 mag when the
spectrum is described with a simple power law ($f(\nu)\propto \nu^p$) of
the index $-2.0<p<0.3$, which is consistent with observed spectra of GRB
afterglows (\cite{sar98grb}).   Our magnitude system was translated to
the $R_{\rm c}$-system by taking cross-correlation with other
observations reported in GRB Coordinate Network (GCN).  The resulting
light curve and magnitudes are shown in figure \ref{fig:lc} and table
\ref{tab:mag}.  In figure \ref{fig:lc}, the abscissa and the ordinate
denote the time from the burst in day and the $R_{\rm c}$-magnitude,
respectively.  Our observations are shown with the filled circles.
Other symbols denote observations reported in GCN.  The exposure time of
each frame was 30 s, and the filled circles denote averaged points of
several frames in the bin of $\Delta \log{t_{\rm day}} \sim 0.07$.
The reason why the first four points have relatively large error bars is
due to low transparency of the sky and shorter total exposure times of
averaged data.  Even in these early points, the afterglow was actually
detected at 4-$\sigma$ level, which is high enough for the discussion in
the text.

\begin{table}
\caption{$R_{\rm c}$-magnitudes obtained by our observation}
\label{tab:mag}
\begin{center}
\begin{tabular}{ccc}
\hline \hline
T (day)$^*$ & $R_{\rm c}$ mag. & err.\\
\hline
0.026568 & 16.86 & 0.53\\
0.031431 & 16.45 & 0.35\\
0.037184 & 16.36 & 0.38\\
0.043990 & 16.60 & 0.38\\
0.072837 & 16.37 & 0.19\\
0.086169 & 16.50 & 0.07\\
0.101942 & 16.69 & 0.06\\
0.120601 & 16.56 & 0.10\\
0.142676 & 16.76 & 0.07\\
0.168792 & 16.91 & 0.08\\
0.199688 & 17.17 & 0.13\\
0.236239 & 17.26 & 0.23\\
\hline
\end{tabular}
\end{center}
{\footnotesize $^*$ Time from the burst (12:06:13.57 UT).}
\end{table}

We initiated time-series CCD photometric observations of GRB021004 on 
October 4 at 12:41:09 UT, about 0.024 day after the burst.  The earliest
positive observations detected a fading of the source from $6.56\times
10^{-3}$ to $1.19\times 10^{-2}$ day after the burst.  Our observation
covers a subsequent period from $2.44\times 10^{-2}$ to 0.292 day.  As
can be seen in figure 1, the first fading trend was terminated before
our run.  Our observations show that, after the initial fading phase,
the object experienced a plateau phase during which it remained at
almost constant brightness.  The object again drastically changed its
fading rate during our observation, and then entered an ordinary decay
phase.  Our $\chi^2$-fitting with two power laws ($f\propto
t^{-\alpha}$) and a transition time yields the best-fit parameters of a
transition time of $0.13\pm 0.02\;{\rm d}$, $\alpha_1=-0.03\pm
0.12\;(t<0.13\; {\rm d})$, and $\alpha_2=1.20\pm 0.36\;(t>0.13\; {\rm
d})$.  We hence confirm that neither significant fading nor
brightening is detectable in the light curve from $2.44\times 10^{-2}$
to 0.132 day.  Even when we use only the earliest period up to
$4.8\times 10^{-2}$ day, the fitting shows no significant fading or
brightening trend.  The fading rate after the plateau phase ($\alpha_2$)
is in agreement with that calculated from other observations reported in
GCN ($\alpha=1.31\pm 0.03$). This value of $\alpha_2$ during the
ordinary decay phase is standard one among known afterglows.

\section{Discussion}

Our observation first revealed the presence of a plateau phase
preceding the ordinary decay of the afterglow.  According to the fireball
model, the optical afterglow is predicted to start rapid fading when the
observed frequency becomes larger than the typical synchrotron
frequency, which rapidly moves to lower-frequency regions with time.  On
the other hand, during a very early phase when the optical flux is still
dominated by the emission under the typical synchrotron frequency, the
flux is predicted to increase, as $f\propto t^{1/2}$ in a case that the
ambient medium has a constant density distribution (\cite{sar98grb}).
The time when the typical frequency passes through the optical range has 
been theoretically estimated to be $\sim 0.1$ d after the burst with
typical physical parameters for GRB afterglows (\cite{sar99prediction}).
We propose that the plateau phase of GRB021004 corresponds to a part of
such an early phase of the ordinary afterglow from the forward shock.
On the nature of the plateau phase, an alternative scenario may be
possible that it is one of wiggles observed during the late phase
($\gtrsim 1\;{\rm d}$), which are apparently unusual compared with other
GRB afterglows.  However, the observed transition time from the early
plateau to the decay phase is just what is expected from the theoretical
calculation, which favors our scenario for the plateau phase.  The
fading trend cannot actually be described by a simple power-law between
the first fading around 0.01 d and the late afterglow after 0.1 d
(\cite{mal02gcn1645}).  The late wiggles in the afterglow light curve
may have just been overlooked in previous GRBs because of their sparse
sampling.  The increase of the fading rate around 5 days can be
naturally interpreted as a normal break as seen in other GRB afterglows.  

The dashed line shown in figure 1 is the best fitted model light curve
based on our picture for the plateau phase.  We chose the smoothly
broken power-law model ($f(t)=(f_1(t)^{-n}+f_2(t)^{-n})^{-1/n}$ with
$f_i(t)=k_i t^{\alpha_i}$) to describe our observations assuming $\alpha
_1=0.5$ and $\alpha_2=1.31$ (\cite{beu990510}).  The parameter $n$
provides a measure of the smoothness of the transition.  It was
calculated to be $n=0.92\pm 0.34$, indicating a very smooth transition
as expected in theoretical predictions (\cite{gra02spec}).  As can be
seen in figure 1, this model well describes our observation.  The
transition time is calculated to be $0.10\pm 0.02\;{\rm d}$ in this
case, which is in agreement with the result from fitting two power-laws
within errors. 

The initial rapid fading trend around 0.01 d should have another
emission source, since, according to the forward shock model, the
optical afterglow reaches its peak just before the ordinary decay phase.
The first fading can be interpreted to be a tail of the optical flash
which was detected only in one GRB afterglow, GRB990123
(\cite{ake99GRB990123}; \cite{kul99GRB990123}).  It has been proposed
that the optical flash originated from a reverse shock, while the
ordinary afterglow originates from a forward external shock
(\cite{sar99GRB990123}; \cite{mes99reverseshock}).  In the case of
GRB990123, the optical flash was observed until at least 0.01 d after
the burst (\cite{kul99GRB990123}).  As seen in figure 1, the transition
time from the flash to the plateau phase was also 0.01--0.02 d in the
case of GRB021004.

As well as the similar activity around 0.01 d of these two GRBs, the
beginnings of the ordinary decay phase were also observed at similar
times around 0.1 d both in GRB021004 and GRB990123
(\cite{kul99GRB990123}).  These two GRBs hence have quite analogous
light curves of afterglows, except for the plateau phase which were
observed not in GRB990123, but in GRB021004.  Our observation has first
filled in the missing blank between the optical flash and the ordinary
decay.  The similarity of these two GRBs and our interpretation with the
natural evolution of afterglows imply that the early light curve of the
afterglow of GRB021004 is common, most fundamental features for GRB
afterglows.  

On the other hand, it is notable that the first fading, which we propose
to be the optical flash, is more gradual compared with the fading branch
of the optical flash in GRB990123 (\cite{sar99GRB990123}).  The
relatively rapid fading after 0.1 d (in other words, a large $\alpha_2$)
may also be rather atypical for other GRB afterglows (\cite{sar99jets}).
These observational features are possibly inconsistent with the picture
which we propose, however, it is too premature to conclude it because we
have only few dense, early phase samples of the light curve of GRB
afterglows, like GRB021004.  The validity of our proposed scenario will
be evaluated by observations of color variations during the early phase
of afterglows since the color of afterglows should dramatically change
when the typical synchrotron frequency passes an observational band.

\vskip 3mm

This work is partly supported by a grant-in aid (13640239) from the
Japanese Ministry of Education, Culture, Sports, Science and Technology.
Part of this work is supported by a Research Fellowship of the Japan
Society for the Promotion of Science for Young Scientists (MU).


\begin{thebibliography}{}

\bibitem[Akerlof et~al.(1999)]{ake99GRB990123}
  Akerlof, C., et~al.\ 1999, \nat, 398, 400

\bibitem[Beuermann et~al.(1999)]{beu990510}
  Beuermann, K., et~al.\ 1999, \aap, 352, L26

\bibitem[Costa et~al.(1997)]{cos97xrayafterglow}
  Costa, E., et~al.\ 1997, \nat, 387, 783

\bibitem[Fox(2002)]{fox02gcn1564}
  Fox, D.~W.\ 2002, GCN, 1564

\bibitem[Frail et~al.(1997)]{fra97radioafterglow}
  Frail, D.~A., Kulkarni, S.~R., Nicastro, S.~R., Feroci, M., \& Taylor, G.~B.\
  1997, \nat, 389, 261

\bibitem[Frail et~al.(2001)]{fra01standard}
  Frail, D.~A., et~al.\ 2001, \apj, 562, L55

\bibitem[Granot, Sari(2002)]{gra02spec}
  Granot, J. \& Sari, R.\ 2002, \apj, 568, 820

\bibitem[Klebesadel et~al.(1973)]{kle73grboriginal}
  Klebesadel, R.~W., Strong, I.~B., \& Olson, R.~A.\ 1973, \apj, 182, L85

\bibitem[Kulkarni et~al.(1999)]{kul99GRB990123}
  Kulkarni, S.~R., {et~al.}\ 1999, \nat, 398, 389

\bibitem[Malesani et~al.(2002a)]{mal02gcn1607}
  Malesani, D., et~al.\ 2002a, GCN, 1607

\bibitem[Malesani et~al.(2002b)]{mal02gcn1645}
  Malesani, D., {et~al.}\ 2002b, GCN, 1645

\bibitem[Meszaros, Rees(1999)]{mes99reverseshock}
  Meszaros, P. \& Rees, M.~J.\ 1999, \mnras, 306, L39

\bibitem[Rhoads(1999)]{rho99lightcurve}
  Rhoads, J.~E.\ 1999, \apj, 525, 737

\bibitem[Sari, Piran(1999a)]{sar99GRB990123}
  Sari, R. \& Piran, T.\ 1999a, \apj, 517, L109

\bibitem[Sari, Piran(1999b)]{sar99prediction}
  Sari, R. \& Piran, T.\ 1999b, \apj, 520, 641

\bibitem[Sari, Piran, \& Halpern(1999c)]{sar99jets}
  Sari, R., Piran, T., \& Halpern, J.~P.\ 1999c, \apj, 519, L17

\bibitem[Sari et~al.(1998)]{sar98grb}
  Sari, R., Piran, T., \& Narayan, R.\ 1998, \apj, 497, L17

\bibitem[Shirasaki et~al.(2002)]{shi02gcn1565}
  Shirasaki, Y. et~al.\ 2002, GCN, 1565

\bibitem[Stanek et~al.(1999)]{sta99GRB990510}
  Stanek, K.~Z., Garnavich, P.~M., Kaluzny, J., Pych, W., \& Thompson, I.\
  1999, \apj, 522, L39

\bibitem[Waxman(1997a)]{wax97fireball}
  Waxman, E.\ 1997a, \apj, 489, L33

\bibitem[Waxman(1997b)]{wax97afterglow}
  Waxman, E.\ 1997b, \apj, 485, L5

\bibitem[Wijers et~al.(1997)]{wij97GRB970228}
  Wijers, R. A. M.~J., Rees, M.~J., \& Meszaros, P.\ 1997, \mnras, 288, L51

\end{thebibliography}
\end{document}